\documentclass[a4paper]{article}
\usepackage[margin=1in]{geometry}
\usepackage{amsmath} 
\usepackage{booktabs}
\usepackage{thesismacros}
\usepackage{lipsum} 
\usepackage[utf8]{inputenc} % allow utf-8 input
\usepackage[T1]{fontenc}    % use 8-bit T1 fonts
\usepackage{hyperref}       % hyperlinks
\usepackage{url}            % simple URL typesetting
\usepackage{booktabs}       % professional-quality tables
\usepackage{amsfonts}       % blackboard math symbols
\usepackage{nicefrac}       % compact symbols for 1/2, etc.
\usepackage{microtype}      % microtypography
\usepackage{graphicx}
\usepackage{algorithmic}
\usepackage{algorithm}
\usepackage{graphicx, subcaption}
\usepackage{bbold}
\usepackage{color}
\usepackage{indentfirst}
\usepackage{amsthm}
\usepackage{tikz}

\usepackage[
	backend=bibtex,
	style=numeric-comp,
]{biblatex}
\newtheorem{theorem*}{Theorem}

\usepackage[T1]{fontenc}
\usepackage{etoolbox}
 
\makeatletter
\patchcmd{\maketitle}{\@fnsymbol}{\@alph}{}{}  % Footnote numbers from symbols to small letters
\makeatother

\title{Mini-batch Tempered MCMC}

\author{
  Dangna Li \\
  ICME\\
  Stanford University\\
  Stanford,  CA, USA\\
  \texttt{dangna@stanford.edu} 
\and
  Wing H. Wong \\
  Statistics Department\\
  Stanford University\\
  Stanford, CA, USA \\
  \texttt{whwong@stanford.edu} \\
}
\date{}

\begin{document}
\maketitle
\begin{abstract}
 In this paper we propose a general framework of performing MCMC with only a mini-batch of data. We show by estimating the Metropolis-Hastings ratio with only a mini-batch of data, one is essentially sampling from the true posterior raised to a known temperature. We show by experiments that our method, Mini-batch Tempered MCMC (MINT-MCMC), can efficiently explore multiple modes of a posterior distribution. Based on the Equi-Energy sampler  (Kou et al. 2006), we developed a new parallel MCMC algorithm based on the Equi-Energy sampler, which enables efficient sampling from high-dimensional multi-modal posteriors with well separated modes. \end{abstract}
%%%\vspace{-5mm}
\section{Introduction}
% brief review of the MH algorithm
The Metropolis-Hastings (MH) algorithm provides a general recipe to sample from a posterior density function given by $\pi(\theta) \propto \pi_0(\theta)\prod_{i=1}^n p(x_i|\theta)$, where $\pi_0(\theta)$  is the prior distribution\footnote{To simplify notation, we assume $\pi_0(\theta)\propto 1$, i.e. the uniform prior over the parameter domain. All the results in this paper generalize to general priors with minor modifications.} of $\theta$ and $\prod_{i=1}^np(x_i|\theta)$ denotes the joint likelihood over an $i.i.d$ dataset $\mathcal{X} = \{x_1,\dots,x_n\}$. The MH algorithm works by building an ergodic Markov chain with $\pi(\theta)$ as its invariant distribution.  To apply the MH algorithm, one needs the ability 1) to evaluate the ratio of the posterior between two points: $\pi(\theta^\prime)/\pi(\theta)$ and 2) to draw sample from a proposal function $q(\theta^\prime|\theta)$.  Given $\theta_t$ at each iteration, we generate $\theta^\prime\sim q(\cdot|\theta_t)$ and then determine whether to accept it with probability\begin{align*}
	r(\theta_t,\theta^\prime) = & \min\left\{1, \frac{\pi(\theta^\prime)}{\pi(\theta_t)}\frac{q(\theta_t|\theta^\prime)}{q(\theta^\prime|\theta_t)}\right\}  \\
	= &  \min\left\{1, \frac{\pi_0(\theta^\prime)}{\pi_0(\theta_t)}\frac{q(\theta_t|\theta^\prime)}{q(\theta^\prime|\theta_t)}\prod_{i=1}^n\frac{p(x_i|\theta^\prime)}{p(x_i|\theta_t)}\right\}
\end{align*}This ``accept or reject'' step (henceforth referred to as the MH correction step) is essential for the Markov chain to converge to the correct invariant distribution. The choice for $q(\theta^\prime|\theta)$ only affects the efficiency of the algorithm.

% the main motivation
When the dataset is large $(n\gg1)$,  evaluating the ratio of the likelihood function at each iteration of MH can be very expensive, as it requires scanning over the entire dataset.
 Instead, researchers tend to rely on  variational Bayes methods or stochastic optimization methods in which each update moves the state along an stochastic gradient direction of the log posterior. Under convexity assumptions, when the step size decreases to zero in a manner that satisfies the Robbins-Monro condition of divergent sum and convergent sum of squares, these updates are guaranteed to find the global optimum. However, without convexity, one can only expect to get close to a local optimum. In fact, it has been shown that MCMC algorithms can have better performance than optimization based algorithms and variational Bayes methods in many applications, including neural network training, topic modeling, matrix factorization, etc
\cite{patterson2013stochastic}\cite{chen2014stochastic}\cite{chen2016bridging}.

In this paper we address the problem of performing MCMC with only a mini-batch of data. Specifically, we show that if one replaces the expensive MH ratio with a cheap mini-batch estimate, one is essentially sampling from $\pi_T(\theta)$ -- the true posterior raised to a certain temperature. Although the samples are not distributed as the true posterior, they are still quite informative from a learning perspective since $\pi_T(\theta)$ keeps all the modes from $\pi(\theta)$. As we shall see, this ``bias'' actually works in our favor by enabling more efficient transitions of the Markov chain. We provide evidences for this claim by applying our method on training Bayesian logistic model. In fact, even for algorithms designed to sample from the exact posterior, to apply them as an inference tool in practice, it may even be necessary to apply tempering on the target distribution. See \cite{seita2016efficient} for an example.

% summarize the contribution of this work
We highlight our contributions as follows: 
%%\vspace{-2mm}
\begin{enumerate}
	\item Although the relationship between tempering and mini-batching has been noted, (for example, \cite{bardenet2015markov} Section 6.3 observed this relationship using a heuristic acceptance rule with the Barker's acceptance criterion \cite{barker1965monte}), our paper is the first  to establish the asymptotic equivalency between tempering and mini-batching in Metropolis-Hastings. 
	%%\vspace{-2mm}
	\item From a practical standpoint, our algorithm is easy to implement and applicable to a wide range of problems. For instances, we do not rely on 
	 any unknown bounds or correction distributions of the log likelihood function, or a carefully designed adaptive sampling scheme \cite{bardenet2014towards, korattikara2014austerity, seita2016efficient}, or the availability of the gradient of the log density (compared with the ``stochastic gradient based approach'' described in Section \ref{sec:relatedwork}).
	%%\vspace{-2mm}
	\item We do not require the step size to be decreasing. This allows a sampler to take much larger steps, which is essential for a sampling algorithm to explore multiple modes of a posterior distribution (more discussions in Section \ref{sec:relatedwork}).
\end{enumerate}
	%%\vspace{-3mm}
%The rest of this paper is organized as follows:
%After reviewing related work in Section \ref{sec:relatedwork}, we present our algorithm and its analysis in Section \ref{sec:method}, where we establish the link between mini-batching and tempering via an augmentation technique. We present the experimental results in Section \ref{sec:mintexp}. As an extension, in Section \ref{sec:mintee} we propose a new MCMC algorithm based on our algorithm, which is capable of sampling from the true posterior in a high dimension with well separated modes. We conclude in Section \ref{sec:conclusions}.

%%\vspace{-1mm}
\section{Related work}\label{sec:relatedwork}
%%\vspace{-1mm}

Below we discuss some related work on the topic of mini-batching in MCMC by summarizing them into 3 categories. An extensive review of all previous work on performing MCMC with mini-batches of data is beyond the scope of this paper. Readers are referred to \cite{bardenet2015markov} for an extensive review.

%%\vspace{-1mm}
 \textbf{The pseudo-marginal approach.} This approach relies on unbiased estimators of the unnormalized target distribution \cite{andrieu2009pseudo, maclaurin2014firefly}. An important feature of methods in this category is that, under certain assumptions, they generate samples from the \emph{exact} target distribution with only mini-batches of data. Unfortunately, these methods are generally not applicable to most problems of interests due to its restrictive assumptions, e.g. a tight and cheap to compute lower bound on the log likelihood function \cite{maclaurin2014firefly}.

%%\vspace{-1mm}
 \textbf{The test based approach.} A method in this category works by approximating the MH test with mini-batches of data. It starts with a small batch size and increases it until a correct decision is made with certain probability. While showing useful reduction in the overall computation, a common drawback of this category is that the amount of data consumed can vary from one MH step to the next.  In addition, these algorithms typically rely on estimating some unknown quantities of the log likelihood or the log likelihood ratio, which makes the implementation complicated and comprises the computation gain \cite{bardenet2014towards, korattikara2014austerity, seita2016efficient}.
 
%%\vspace{-1mm}
\textbf{The stochastic gradient based approach.}
% the problem of decreasing step size
A typical algorithm of the this category uses the stochastic gradient to construct their proposal and discard the MH correction step altogether (i.e. it always moves to the new proposed point. See, for example, \cite{ahn2012bayesian, welling2011bayesian, chen2014stochastic, chen2016bridging}).
However, without the MH correction step, there is no guarantee that such an algorithm will generate samples from the correct distribution. In fact,  \cite{bardenet2015markov} showed naively approximating the MH test with a random mini-batch leads to a complicated invariant distribution that is hard to interpret; \cite{chen2014stochastic} showed a naive implementation of ``mini-batch Hamiltonian Monte Carlo'' will generate samples that are arbitrarily bad. To make it safe to discard the MH correction step, one common assumption made by algorithms in this category is the ``decreasing step size'' assumption. Intuitively, without a correction step, each iteration is likely to introduce a certain amount of bias. As the biases accumulate over iterations, the Markov chain may not even be converging. This bias can be reduced if one uses a very small step size, such that the landscape of the posterior distribution does not change much between moves. In this case, the acceptance probability of the proposed move will be close to 1, hence making it safe to avoid a correction step. However, besides slowing down the mixing of the chain, a more fundamental problem of the decreasing step size assumption is that it implies the algorithm will converge to a local mode instead of moving between modes with probability consistent with the posterior, at least not within a reasonable time frame.

 It is worth pointing out that our method is related to the usage of fractional posterior, which is obtained by updating a prior distribution with the usual likelihood function raised to a fractional power, the fractional likelihood function. There has recently been an renewed interests in the usage of fractional posterior in Bayesian statistics, largely due to its empirically demonstrated robustness to model misspecification. Moreover, \cite{bhattacharya2016bayesian} provide theoretical analysis on the contraction property of the fractional posterior in a general misspecified setting and develop oracle inequalities for the fractional posterior. As we shall see, our method is asymptotically equivalent to the use of fractional posterior, since the joint likelihood will dominate the prior when the data set is sufficiently large. 
%%\vspace{-1mm}
%%\vspace{-1mm}
\section{Mini-batch Tempered MCMC}\label{sec:method}
%%\vspace{-1mm}
\subsection{The algorithm}
%%\vspace{-1mm}
Before delving into the detailed analysis, we first present our algorithm: MINT-MCMC(``MINi-batch Tempered MCMC''), or MINT for short. Let  $\mu(\theta) = \frac{1}{n}\sum_{i=1}^n l_i(\theta)$, where $l_i(\theta)\equiv l(x_i;\theta) = \log p(x_i|\theta)$, i.e. the log likelihood of data point $x_i$ evaluated at parameter $\theta$. Let  $\hat{\mu}(\theta) = \frac{1}{m}\sum_{j=1}^ml_{i_j}(\theta)$, where $\{i_1,\dots, i_m\}$ is a random subset sampled uniformly with without replacement from $ \{1,\dots, n\}$. In other words, $\hat{\mu}(\theta)$ is an estimate of $\mu(\theta)$ based on a mini-batch of size $m$. Our algorithm can be described in pseudo-code as in Algorithm \ref{algo:mint}. It has two important input parameters: $\tau$, which determines the batch size; and $\lambda (\lambda < \tau)$ which determines the temperature of the approximate stationary distribution.

We would like to highlight an important difference between our algorithm and the original MH algorithm: the scaling factor for log likelihood difference is $n^{\lambda}$ instead of $n$. It has been noted in \cite{bardenet2015markov} that naively using a mini-batch estimate for the log likelihood can lead very poor results, i.e. estimating $\mu(\theta)$ by $\frac{n}{m}\hat\mu(\theta)$. Importantly, our work suggests that the cause of the failure of the naive approach is the incorrect scaling factor. 

\begin{algorithm}[htbp]                   % enter the algorithm environment
\caption{MINT}
\label{algo:mint}                           % and a label for \ref{} commands later in the document
\begin{algorithmic}[1]                    % enter the algorithmic environment

    \REQUIRE $q(\theta\rightarrow \theta^\prime)$, $\tau$, $\lambda(< \tau)$, $l(x;\theta)$ and $B$ (number of burn-in samples)
     \ENSURE $\theta_{B+1},\dots,\theta_{N}\sim\pi_T(\theta)$
    \STATE $t = 0$

    \WHILE{$t \leq Nan$}
    	\STATE $\theta = \theta_t$
    	\STATE Propose a move $\theta^\prime $ using $ q(\theta\rightarrow\theta^\prime)$
    	\STATE Compute the mini-batch MH ratio: $$r = \min\left\{e^{{n^\lambda(\hat\mu(\theta^\prime)-\hat\mu(\theta))}}\frac{q(\theta^\prime\rightarrow\theta)}{q(\theta\rightarrow\theta^\prime)},1\right\}$$
	\STATE Draw $u$ uniformly from $[0,1]$
        \IF{$u < r$ }
            \STATE $\theta_{t+1} = \theta^\prime$
        \ELSE
            \STATE $\theta_{t+1} = \theta$
        \ENDIF
        \STATE $t = t + 1$
    \ENDWHILE
\end{algorithmic}
\end{algorithm}
%%\vspace{-2mm}
\subsection{Analysis}\label{sec:analysis}
%%\vspace{-2mm}
In this section we prove the following result: suppose the mini-batch size is chosen as $m = n^{\tau}$ where $0<\tau<1$,  $\lambda<\tau$, then the MINT simulates a Markov chain whose invariant distribution is asymptotically the true target distribution raised to the temperature $T = n^{1-\lambda}.$

%%\vspace{-1mm}
We analyze MINT with an augmentation technique.  For reasons that will become clear soon, we consider first sampling from an augmented system $(\theta,t)\in\Omega\times\mathcal{R}$, whose joint density function is given by $f(\theta,t)\propto g(\theta)e^{\epsilon t}\phi_\theta(t)$, where $g(\cdot)$ is some density function of $\theta$; $\phi_\theta(\cdot)$ is the probability density function of $\Gsn{(0, \sigma^2_\theta)}$ whose variance may depend on $\theta$; $\epsilon$ is some constant which does not depend on  $\theta$ or $t$. We can sample from this new system using the MH algorithm with proposal $q((\theta,t)\rightarrow (\theta^\prime,t^\prime)) = q(\theta\rightarrow\theta^\prime)\phi_{\theta^\prime}(t^\prime)$, where $q(\theta\rightarrow\theta^\prime)$ is some arbitrary proposal function of $\theta$, e.g. a random walk centered at $\theta$.  This defines a homogeneous Markov chain in the augmented space. From standard Markov chain theory we know that $f(\theta, t)$ is the invariant distribution of this chain. The marginal distribution of $\theta$ of this chain is given by: $$
f^*(\theta) = \int_\reals f(\theta, t)dt\propto  g(\theta)\int_\reals e^{\epsilon t}\phi_\theta(t)dt = g(\theta)e^{\frac{1}{2}\sigma_\theta^2\epsilon^2}
$$where the last equality follows from the moment generating function of $\Gsn{(0, \sigma^2_\theta)}$. 

%%\vspace{-1mm}
We now apply the above augmentation idea to derive our algorithm.  Write%%\vspace{-2mm} 
\begin{equation}\label{def:t}
	t = \sqrt{m}(\hat{\mu}(\theta) - \mu(\theta))
\end{equation}
%%\vspace{-1mm}
%\begin{equation}\label{def:sigma}
%	\sigma^2_\theta = \frac{1}{n}\sum_{i=1}^n(l_i(\theta) - \mu(\theta))^2
%\end{equation} 

Then under random mini-batching, by the Central Limit Theorem, $t\sim\Gsn{\left(0,\sigma^2_\theta\right)}$ with a high degree of accuracy, where $\sigma_\theta^2$ denotes the variance of $l_i(\theta)$. Let $m=n^{\tau}$ where $\tau\in[0, 1)$ is a hyper-parameter for batch size. Define $\tilde{\pi}(\theta,t) = e^{n^\lambda\hat{\mu}(\theta)}= e^{n^\lambda \mu(\theta) + n^{\lambda-\tau/2}t}$, where $\lambda<\tau$ is another hyper-parameter for temperature. Notice $\tilde{\pi}(\theta,t)$ can be evaluated using only a mini-batch of data. Now let $n^{\lambda - \tau/2} = \epsilon$, $g(\theta)\propto e^{n^\lambda \mu(\theta) }$, consider the following joint distribution of $(\theta ,t)$:
\begin{equation}
	f(\theta,t)\propto g(\theta)e^{\epsilon t}\phi_\theta(t)\equiv \tilde{\pi}(\theta,t)\phi_\theta(t)
\end{equation}From our discussion at the beginning of this section, if we assume normality for $t$, we can integrate it out to obtain: $$f^*(\theta) \propto e^{n^\lambda \mu(\theta) + \frac{1}{2}\sigma^2_\theta\epsilon^2} =  e^{n^\lambda \mu(\theta)\left(1 + n^{-(\tau - \lambda)}\gamma(\theta)\right)} $$ as the marginal equilibrium distribution for $\theta$, where  $\lambda < \tau$, $\gamma(\theta) = \frac{\sigma_\theta^2}{2\mu(\theta)}$. If $\gamma(\theta)$ is uniformly bounded, then  as $n\rightarrow\infty$ this marginal is approximately the tempered posterior with $T = n^{1-\lambda}$.

   %%\vspace{-1mm}
 Therefore, if we can sample from $f(\theta, t)$, the generated samples of $\theta$ will be marginally distributed as the true posterior raised to a temperature of $n^{1 -\lambda}$. We now show this can be achieved by touching a mini-batch of data at each iteration. Recall to sample from the joint distribution $f(\theta, t)$, we can use the MH algorithm with proposal $q((\theta,t)\rightarrow (\theta^\prime,t^\prime)) = q(\theta\rightarrow\theta^\prime)\phi_{\theta^\prime}(t^\prime)$. On the first glance it may seen inevitable to touch all $n$ points in the dataset, since the variance of $t$, $\sigma^2_\theta$, is usually unknown and its unbiased estimate $s_\theta^2=\frac{1}{n-1}\sum_{i=1}^n(l_i(\theta) - \mu(\theta))^2
$ requires a scan over the entire dataset. However, it turns out that  to compute the Metropolis-Hastings ratio, knowledge of $\sigma^2_\theta$ is not required: \begin{align*}
\frac{f(\theta^\prime,t^\prime)q( (\theta^\prime,t^\prime)\rightarrow (\theta,t))}{f(\theta,t)q((\theta,t) \rightarrow (\theta^\prime,t^\prime))} 
  =& \frac{\tilde{\pi}(\theta^\prime,t^\prime)\phi_{\theta^\prime}(t^\prime)q(\theta^\prime\rightarrow\theta)\phi_{\theta}(t)}{\tilde{\pi}(\theta,t)\phi_\theta(t)q(\theta\rightarrow\theta^\prime)\phi_{\theta^\prime}(t^\prime)} \\
=& 
 \frac{\tilde{\pi}(\theta^\prime,t^\prime)q(\theta^\prime\rightarrow\theta)}{\tilde{\pi}(\theta,t)q(\theta\rightarrow\theta^\prime)} 
 \end{align*}
 
 %%\vspace{-1mm}
 
  That is, the terms involving $\sigma_\theta^2$ cancel out.
In other words, if all we care about is $\theta$, we can construct an implicit Markov chain in the augmented space by using only a mini-batch of data at each MH step. The stationary distribution of this Markov chain is the true posterior raised to a temperature of $n^{1-\lambda}$.

% notations! 

%The objective for our algorithm is not to sample from the true posterior exactly, but to utilize the efficiency of MINT to explore the parameter space efficiently. This is especially attractive when the posterior density function is non-convex. Since the invariant distribution of MINT is a ``tempered'' version of the true posterior, it inherents all the modes from the  true posterior. This means we can use our algorithm as a first stage algorithm to locate a rough position of multiple modes and then use some optimization routine to zoom in on the parameter space and locate the modes exactly. Compared with optimization based method, our sampling based method is less likely to get stuck in local optima and more likely find a better local optima, if not the global optima.
%%%%\vspace{-1mm}
\begin{figure*}
\centering
\includegraphics[width=.9\textwidth]{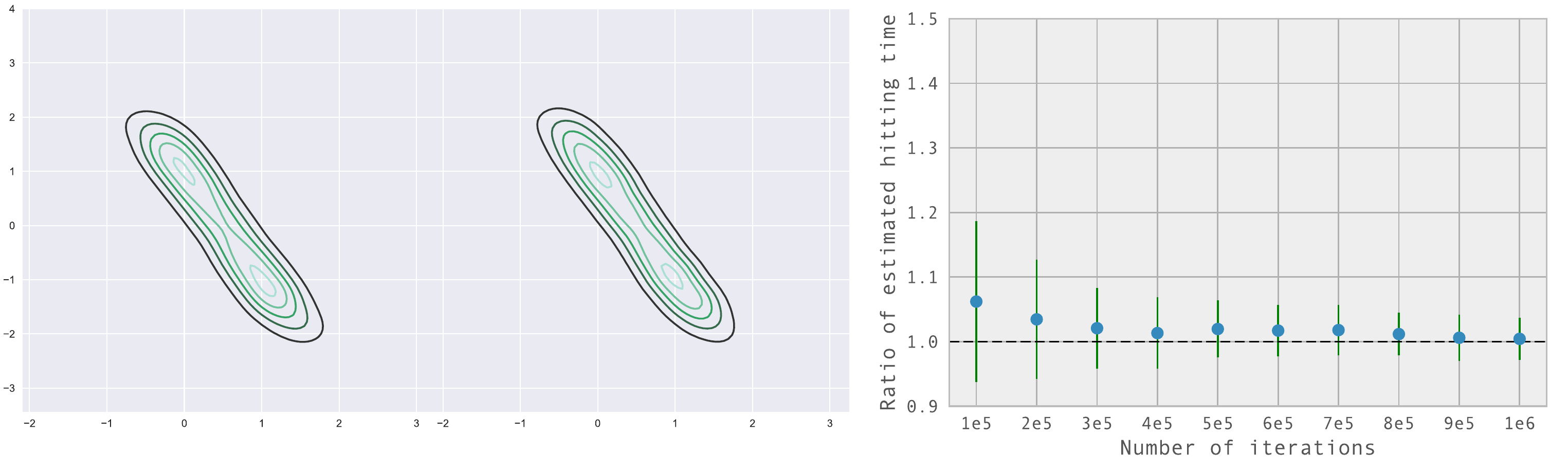}
%%\vspace{-4mm}
\caption{Experimental results of mixture of Gaussians with tied means. \emph{left}: contour plot of samples of MINT and tempered MCMC. \emph{right}: average hitting time estimated from 20 independent runs as a function of the number of iterations.  The green bars indicate 1 standard error bars. The black dashed line represents the ground truth. }
%%\vspace{-2mm}
\label{fig:sgld}
\end{figure*}

\subsection{MINT in practice}%%\vspace{-1mm}
Compared with standard MH algorithms, MINT only needs two additional parameters, $\tau$ and $\lambda$, where $\tau$ is a parameter to control the batch size $m=n^{\tau}$. $\lambda$ is a parameter to control the temperature of the invariant distribution of the chain. A necessary condition for $\lambda$ is $\lambda < \tau$. Under this constraint, the higher the $\lambda$, the lower the temperature, which implies the invariant distribution will be closer to the true posterior. If we choose a small $\lambda$, then the invariant distribution will be more flatten out, which will facilitate movement between modes. As a general guideline, if one's goal  is locating modes of the posterior, it is desirable to choose a larger $\lambda$ (say, $\lambda = 0.99\tau$). On the other hand, if the goal is to transition between modes, it is worthwhile to consider a smaller $\lambda$. We will illustrate in Section \ref{sec:mintexp} the choice of $\lambda$ with several concrete examples.

%%\vspace{-2mm}
\section{Experiments: sampling with MINT}\label{sec:mintexp}%%\vspace{-1mm}
In the rest of this paper, we denote $\pi_T(\cdot)$ as the true posterior raised to temperature $T$, with the understanding that $T = n^{1-\lambda}$. 
We use the term ``full-batch MCMC'' to refer to traditional MCMC algorithms which use the entire dataset to estimate the MH ratio. We refer to full-batch MCMC applied on $\pi_T(\theta)$ as ``tempered MCMC''. The term ``MCMC'' is reserved for full-batch MCMC applied on the true posterior.
Unless otherwise specified, we use Gaussian random walk proposals with a constant step size for an experiment. The step size of a proposal is chosen such that the acceptance probability of a sampler is around $0.30$. After the  batch size $m$ is chosen, we specify our parameter choice through $\lambda = \alpha\tau$, $0<\alpha<1$, where $\tau$ can be deduced from the batch size via $\tau =\log m/\log n$.
%%\vspace{-2mm}
\subsection{Gaussian mixture with tied means}
%%\vspace{-1mm}
We first demonstrate the workings of MINT on a simple example. To make the posterior multimodal, we use the example of  mixture of Gaussians with tied means as in \cite{welling2011bayesian}:
%%\vspace{-1mm}
\begin{align*}
	& \theta_1\sim \Gsn{(0, \sigma_1^2)},\quad \theta_2\sim \Gsn{(0, \sigma_2^2)}\\
 & x_i\sim \frac{1}{2}\Gsn{(\theta_1,\sigma_x^2)} + \frac{1}{2}\Gsn{(\theta_1+\theta_2,\sigma_x^2)}\end{align*}We adopt a similar parameter setting as in \cite{welling2011bayesian}: $\sigma_1^2=10, \sigma_2^2=1, \sigma_x^2=2$, except for a much larger sample size: we draw $10^6$ observations from the model with $\theta_1 = 0$ and $\theta_2=1$. Besides one mode at this point, the posterior has another mode at $\theta_1=1$, $\theta_2=-1$. Due to the symmetry in the model, the ratio of the true posterior between this two modes is 1.00.  We test MINT on this example by generating $10^6$ samples with a batch size of $1,000$ and $\lambda = 0.5\tau $. \mbox{The results are summarized in Figure \ref{fig:sgld}.}

The first two plots in Figure \ref{fig:sgld} demonstrate the closeness of the empirical distribution of samples generated by MINT and $\pi_T(\theta)$, respectively. In addition, it shows MINT captures both modes accurately. In the right panel of the same figure, we test MINT by estimating the ratio of the true posterior between these two modes.  We estimated the ratio as the ratio between the empirical probabilities for the samples to fall within a small spherical neighborhood of radius $1e^{-2}$ of a mode. For a fixed number of iterations, we estimate this ratio by taking an average over 20 independent runs. The results show MINT can estimate this ratio quite accurately. We also benchmarked our algorithm against several other algorithms: Stochastic Gradient Langevin Dynamics (SGLD) \cite{welling2011bayesian} and Stochastic Gradient Hamiltonian Dynamics (SGHMC) \cite{chen2014stochastic}. These two algorithms were not able to  provide a valid estimate for the ratio within the $10^6$ iterations, since they can only visit one of the two modes, depending on the starting position. As noted previously, this is because they both require the step size to be annealed to zero and hence making it difficult for them to escape a local mode.

\subsection{Bayesian logistic regression}
\begin{figure}
\centering 
\includegraphics[width=\linewidth]{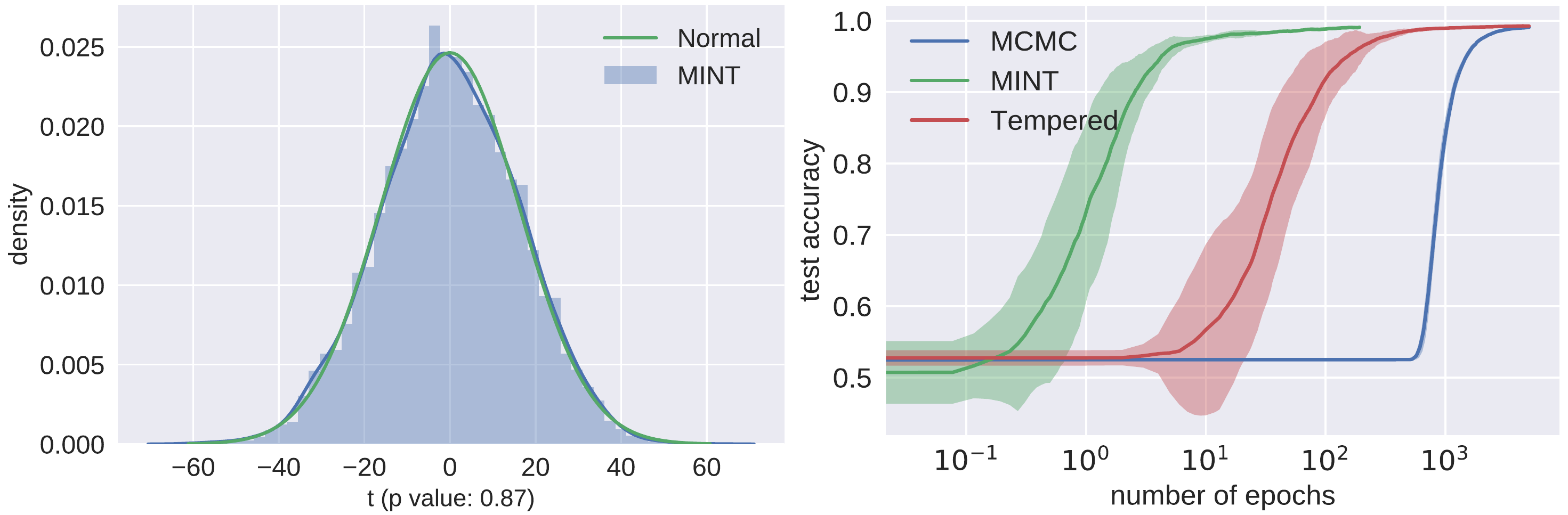}
%%\vspace{-3mm}
\caption{Experimental results of Bayesian logistic regression. \emph{left:} comparison of the empirical of
distribution $t$ with a univariate gaussian with the same mean and variance. \emph{right:} convergence of test accuracies for MINT, tempered MCMC and MCMC. The shaded regions indicate standard errors over 10 independent runs. 
Here 1/130 epoch is equivalent to one mini-batch iteration. }
 \label{fig:logreg}
%%\vspace{-6mm}
\end{figure}
%%\vspace{-1mm}
Next, we test our algorithm on training Bayesian logistic regression models for 1s versus 7s binary classification task on the MNIST \cite{lecun2010mnist} dataset which includes $28\times 28$ gray scale images for hand-written digits. The data includes 13,000 training samples and 2,163 test samples after extracting all the 1s and 7s.

%%\vspace{-1mm}
Recall that the only approximation we made in the derivation in Section \ref{sec:method} is the normality of $t$:
%%\vspace{-1mm}
\begin{equation*}
	t = \sqrt{m}(\hat{\mu}(\theta) - \mu(\theta))\sim\Gsn{\left(0,\sigma^2_\theta\right)}
\end{equation*}To investigate the validity of this assumption, for a fixed $\theta$, we draw 5,000 random batches of size 100 from the training data and estimate the distribution of $t$. In the left panel in Figure \ref{fig:logreg}, we compare the empirical distribution of $t$  against a univariate Gaussian with the same mean and variance. The comparison shows the normal approximation is quite close for this example. 

%%\vspace{-1mm}
Next, we compare MINT with  MCMC and tempered MCMC in terms of classification accuracy.In the right panel of Figure \ref{fig:logreg}, we plot the test accuracy of different samplers against the number of passes made through the training data. Since full-batch MCMC algorithms need to make an entire pass over the training data to generate one sample, MCMC and tempered MCMC did not start to make progress until after 1 epoch\footnote{In this paper we use the word ``epoch'' and ``an entire pass over the training data'' interchangeably.} and their accuracy did not converge until after $100$ epochs. In comparison, MINT is able to generate $n/m=130$ samples each pass over the training set (In the figure, 1/130 epoch is equivalent to one mini-batch iteration). In this case, the result shows the accuracy of MINT converges a several magnitude faster than tempered MCMC and MCMC. This suggests that in terms of “number of bits learned per unit of computation”, MINT is a lot more efficient than its full-batch counterparts. In the appendix, we include a similar plot but with wall clock time measured in seconds on the $x$-axis. There we see MINT achieves the same accuracy with a 100-fold speed-up over MCMC. 
\begin{table}
 %%\vspace{-4mm}
  \caption{Average mini-batch sizes($\pm$ one standard deviation) on logistic regression on the MNIST dataset. 
The averages are taken over 10 independent runs (5000 samples each). ``**'' indicates results  taken from \cite{seita2016efficient}. }
  \label{tab:test_based}
  \centering
  \begin{tabular}{c|c}
    \toprule
Method & Mini-batch size\\
    \midrule
    MINT & 100 \\
MHMINIBATCH** &  $125.4 \pm 9.2$ \\
AUSTEREMH(NC)** & $973.8 \pm 49.8$ \\
AUSTEREMH(C)** & $1924.3 \pm 52.4$ \\
MHSUBLHD** & $10783.4 \pm  78.9 $ \\
\bottomrule
  \end{tabular}
 %%\vspace{-7mm}
\end{table}

Another interesting observation from the results is that tempered MCMC seems to converge much faster than MCMC on the true posterior. This is because the tempered posterior is much flatter then the true posterior, which in this example is highly concentrated around its maximum due the size of the training set. Since we keep the acceptance probabilities of different samplers to be roughly the same, tempered MCMC is able to take much larger steps.  In fact, among the accepted transitions, the average step sizes of MINT, tempered MCMC and MCMC turned out to be 1.71, 1.50 and $0.05$, respectively. Given the asymptotical equivalency of MINT and tempered MCMC, this in turn explains the efficiency of MINT from an exploration perspective.
% [todo] compare with other test based method on this example

In Table \ref{tab:test_based} we compare the batch size used by MINT with several other test based mini-batch MCMC algorithms, including \cite{bardenet2014towards}(MDSUBLHD), \cite{korattikara2014austerity}(both the non-conservative version AUSTEREMH(NC) and the conservative version AUSTEREMH(C) ) and \cite{seita2016efficient}(MHMINIBATCH).  Given their probabilistic nature, the batch sizes they use per iteration can vary, where as MINT uses a constant batch size. In fact, all these methods achieve comparable test accuracy ($>99\%$) on this example. However, in terms of mini-batch size, MINT not only uses the smallest batch size but is the only method whose batch size remains constant.

It is important to point out an important difference between MINT and these other algorithms: the goal of these test based methods is to sample from the true posterior while MINT samples from a tempered posterior. However, even if an algorithm was original designed to sample from the true posterior, it may be unavoidable to increase the temperature of the target posterior to facilitate parameter exploration. See \cite{seita2016efficient} for an example.

\section{Mini-batch Equi-Energy Sampler}\label{sec:mintee}
%%\vspace{-1mm}
\subsection{Background}\label{sec:MINTEE:bg}
%%\vspace{-1mm}
When the posterior is multimodal, MCMC methods will offer more accurate inference than that based on point estimation such as MAP (maximum a posteriori) or MLE (maximum likelihood estimate). On the other hand, it is well-known that when the posterior is multimodal, a naive MCMC sampler is likely to be trapped in one of the modal region. As a result, the sampler will miss, or vastly underestimate the probabilities of, the other modal regions. When the sampler is trapped near a local mode, the chain can be mixing so slowly that it may appear to have reached equilibrium when this is in fact can be quite misleading. This problem of slow mixing is a main reason a MCMC based inference algorithm fails to deliver good results.

%%\vspace{-1mm}
To handle this problem, in 1991, Charles Geyer introduced ``parallel tempering'' , which to this date is still one of the most effective methods to achieve fast mixing. The key idea is to sample from a sequence of tempered distributions. Specifically, the algorithm considers a temperature ladder $1=T_0<T_1<T_2<\cdots<T_{K-1} < \infty$ and  evolves $K$ Markov chains in parallel.  Each of the $K$ chains samples from a tempered distribution at a different temperature $T_k$,  $k=0,\dots, K-1$.  With a slight abuse of notation, denote by $\pi_k(\cdot)$ the tempered version of $\pi(\cdot)$ at temperature $T_k$. It constructs a Markov chain in the joint space $(\theta^{(0)}, \theta^{(1)}, \cdots, \theta^{(K-1)})$, where $\theta^{(k)}\sim\pi_k(\cdot)$, $k=0,\dots K-1$. Occasionally, a proposal is made to exchange the states of two chains. The proposal is accepted or rejected according to a MH rule designed to sample from the product of the $K$ tempered densities $\pi_0(\theta)\pi_1(\theta)\cdots\pi_{K-1}(\theta)$. In other words, each of the $K$ chains is evolved with two types of transitions: Type I) Update $\theta^{(k)}$ by a MH move  designed to sample from $\pi_k(\theta)$, with the chains at different temperatures updating independently of each other; Type II) Exchange current values of $\theta^{(k)}$ and $\theta^{(k+1)}$. Here the MH ratio based on the product joint density is used to determine the acceptance of the proposal.  When $T$ is high, the landscape between the modes will be flattened out. This makes it easier for a MCMC sampler to escape from local modes and thereby speed up the mixing of a single chain. By the exchange moves, fast mixing at high temperatures will accelerate mixing at lower temperatures. As a result, the whole system will mix at a much faster rate than that of a single chain at $T =1$ evolving by itself. See \cite{earl2005parallel} for a review of applications of parallel tempering in hard simulation problems.

%%\vspace{-1mm}
In addition to parallel tempering, multi-canonical sampling \cite{berg1992multicanonical} via the Wang-Landau algorithm is another powerful method for simulating physical systems with complex energy landscape. Later, \cite{kou2006discussion} combines key elements from multi-canonical sampling and parallel tempering with a new type of movement, the Equi-Energy jump, to form the ``Equi-Energy  sampler (EE sampler)''. More details of the EE sampler will be discussed in Section \ref{sec:MINTEE:method}. Here we note that fast mixing in all of these advanced methods is due to sampling from a sequence of highly tempered distributions. However, when $\pi(\cdot)$ is a posterior density given $n$ i.i.d observations, it remains an open challenge on how to sample from a tempered $\pi(\theta)$.
%%\vspace{-1mm}
%%\vspace{-1mm}
%%\vspace{-1mm}
\subsection{The MINTEE sampler}\label{sec:MINTEE:method}
%%\vspace{-1mm}
%%\vspace{-1mm}
As an extension to MINT, in this section we introduce a new sampling algorithm based on MINT and the EE sampler: ``MINi-batch Tempered Equi-Energy  sampler (MINTEE)''. The idea is straight forward: we apply MINT to sample from a series of tempered posteriors and incorporate EE jump to exchange information between chains. The mixing of a low temperature chain will be accelerated by the fast mixing of a high temperature chain. Although our focus here is to combine MINT with the EE sampler, the combination with parallel tempering is completely analogous. 

%%\vspace{-1mm}
%%% How does EE work %%%
Similar with parallel tempering, the EE sampler considers a temperature ladder: $ 1 = T_0 < T_1 < \cdots T_{K-1} <\infty
$. In addition, it also considers an increasing energy sequence:
$ H_0<H_1<H_2<\cdots H_{K-1} < H_{K} = \infty$ where $H_0$ is a lower bound for the energy function: $H_0\leq\inf_xh(x)$. This defines $K$ distributions, each indexed by a temperature and an energy truncation: $\pi_k(\theta) \propto \exp{- \frac{h(\theta)\vee H_k}{T_k}}$ for $k=0,1,\cdots K-1$. For each $k$, a sampling chain targeting $\pi_k(\cdot)$ is constructed.  Clearly, $\pi_0(\cdot)$ is the original target distribution. The EE sampler employs the other $K-1$ chains to overcome local trapping by incorporating a type of transition called the ``Equi-Energy  jump (EE jump)'' to exchange states between two adjacent chains. In the appendix, we include a detailed description of the EE sampler, together with a pictorial illustration and a comparison with parallel tempering. Below we will focus on discussions unique to MINTEE.

%%\vspace{-1mm}
To combine MINT with the EE Sampler, we propose to use MINT to sample in parallel the $K-1$ tempered truncated distributions, $\pi_i(\theta), i=1,\dots K-1$. We choose a small starting batch size $m_{K-1}$, e.g. $m_{K-1} = 1000$ when $n = 1e6$, and increase the batch size geometrically along the ladder: $m_{k} = \gamma m_{k+1}$, $\gamma > 1$. The temperature of the chain is implicitly given by the relations: $\tau = \log m/ \log n$, $T = n^{1-\lambda}, \lambda < \tau$. In practice, typically one would want to choose $\lambda$ close to $\tau$ so that the tempered distributions will be closer to the true target distribution. It should be noted the maximum length of the ladder is bounded by the sample size $n$, \mbox{the smallest batch size $m_{K-1}$ and its rate of increase $\gamma.$}

%%\vspace{-1mm}
A unique advantage of MINTEE over the EE sampler is its computational efficiency. Suppose we want to collect $N$ samples from a posterior of $n$ $i.i.d$ observations with $K$ chains. Without MINT, the total cost of applying the EE sampler is $O(K^2Nn)$, whereas the overall cost of MINTEE can be easily shown to be $O(Nn)$, assuming the smallest batch size  $m_{K-1}$ is chosen such that $m_{K-1}\gamma^{K-1} = n$.

%We start with sampling from the highest temperature chain by running MINT with batch size $m_K$. After this chain appeared to have reached equilibrium, we start the next lower temperature chain by setting $m_{k} = \gamma m_{k+1}$ and run MINT with the new batch size. After this chain has reached equilibrium, we start another chain with a even lower temperature, etc. Fin

%%\vspace{-2mm}
\subsection{Experiments}
%%\vspace{-2mm}
\begin{figure}
\centering
\includegraphics[width=0.9\textwidth]{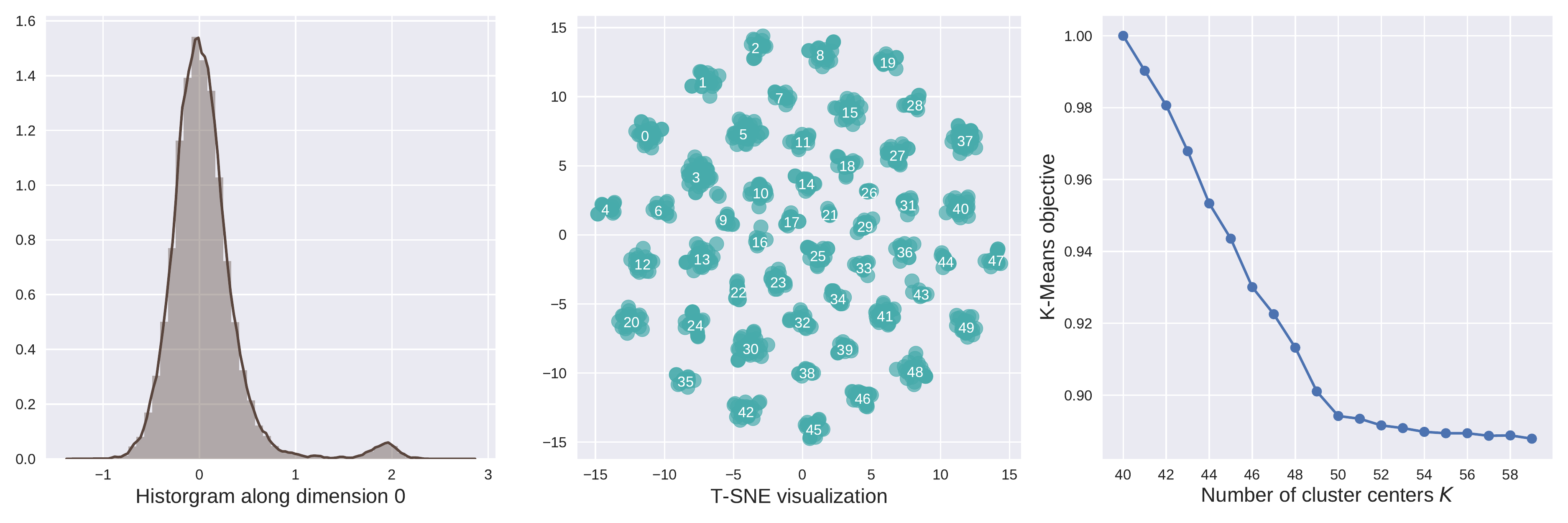}
%%\vspace*{-3mm}
	\caption{Visualizations of samples from the lowest temperature chain. \emph{left}: 1-D histogram along the first dimension. \emph{middle}: scatter plot of 2d T-SNE embedding. We labeled each ``cluster'' in the 2d embedding with a numerical label to aid visualization. \emph{right}: plot of $k$-means objective value against number of clusters. }
	
		\label{fig:MINTEE:50d}
	%%\vspace{-6mm}
\end{figure}
We include the pseudo code and implementation details of MINTEE in the appendix . As an illustration, consider the following model for $x\in\reals$: $x\sim \frac{1}{50}\Gsn(\theta_1, \sigma^2) + \frac{1}{50}\Gsn(\theta_2, \sigma^2) + \cdots + \frac{1}{50}\Gsn(\theta_{50}, \sigma^2)$ where $\sigma^2=1$ is assumed to be known; $\theta=(\theta_1, \theta_2,\cdots,\theta_{50})\in\reals^{50}$ is the parameter of interests. We generate $10^5$ observations from $\theta^* = [2, 0, \cdots, 0]$. Due to the symmetry of the model in $\theta$, the posterior has 50 modes locating at the corners of a hyper-rectangular of size $2^{50}$. This posterior serves as a good test for sampling algorithms since the dimension is relatively high and the modes are quite far away.

%%\vspace{-1mm}
We run MINTEE to generate 300,000 samples from the lowest temperature chain. More implementation details can be found in the appendix, where we also include results showing samples from MINTEE follow the correct marginal distribution. In Figure \ref{fig:MINTEE:50d}, we visualize the  the results for the lowest temperature chain. In the left panel we plot the 1 dimensional histogram of the first dimension of the samples. The result shows clearly that the samples have the correct marginal distribution: the samples follow a bimodal distribution whose the height of the mode at $\theta_1=0$ is about 50 times of mode at $\theta_1=2$, which is consistent with the posterior. We made the same plot for other 49 dimensions and found similar results. It is worth noting that in this case any optimization method will only be able to locate one of the 50 modes, failing to provide any information about the relative probability   of the local optima.  We plot in the middle panel a 2 dimensional embedding learned by T-SNE \cite{maaten2008visualizing}. The plot shows MINTEE has visited all 50 modes with roughly equal probabilities. Next, we examined the clustering properties of the sampling using the elbow rule of K-Means. The plot shows a clear  elbow shaped curve with a turning point at $C=50$, which shows the samples are highly concentrated around the 50 modes. We also compared the sampling results of MINTEE with plain MCMC. In this example, MCMC failed to escape from the local mode where it starts. 
%\vspace{-3mm}
\section{Conclusions and future work}
\label{sec:conclusions}
%%\vspace{-2mm}
In this work we propose a mini-batch  MCMC algorithm by establishing the asymptotic equivalency between mini-batching and tempering. Our method builds on the fundamental framework of traditional MCMC, but uses stochastic estimate of the MH ratio to avoid the costly exact computation. By using an augmentation technique, we prove the samples generated by MINT is asymptotically distributed as the true posterior raise to a known temperature. In addition, our derivation shows the root cause to the failure of the naive subsampling approach \cite{chen2014stochastic, bardenet2015markov} is the incorrect scaling factor. Our empirical results, both in simulated settings and on real data, validate our theory and demonstrate the practical value of MINT as an inference tool. Based on MINT, we also developed MINTEE, which is a sampling algorithm which can sample exactly from a high dimensional  posterior with well-separated modes.

In terms of future work, it would be interesting to extend MINT to methods such as Hamiltonian Monte Carlo (HMC). \cite{betancourt2015fundamental} shows the sample trajectories generated by naive mini-batching can deviate from the trajectories generated using full data, unless the batch size is so large that rules out any computational gain. \cite{chen2014stochastic} also shows naively doing subsampling in HMC can break any convergence guarantees to the true distribution. However, as we have emphasized in the paper, MINT's convergence properties are ensured by the use of a mini-batch MH step which is independent of the proposal function. It thus seems promising that one should be able to incorporate HMC within MINT's framework. 

Another direction worth pursuing is to consider how to perform mini-batch Gibbs sampling. As early as 2004, \cite{rosen2004author} provides empirical evidence showing applying mini-batching within Gibbs sampling for Latent Dirichlet Allocation \cite{blei2003latent} delivers excellent results.  More recently,  \cite{johndrow2015approximations} provides insights for doing mini-batching in Gibbs sampling on a large scale logistic regression, by tackling the problem from a more general perspective of approximating the Markov transition kernel. Given existing work so far, we believe one should be able to generalize the idea we used in developing MINT to Gibbs sampling.  

On the theory side, it would be worthwhile to study the theoretical properties of MINT is greater depth. For example, it would be insightful to study the rate of convergence of MINT to the tempered distribution. However, they are beyond the scope of this paper and we leave them for future work.

\section*{Acknowledgement}

We are grateful to James Johndrow for his useful comments on the paper. We thank Rachel Wang and Tung-yu Wu for the helpful discussions during the early stage of this project. The work is supported by NIH-R01GM109836 and NSF-DMS1407557.

\newpage
\nocite{*}
\printbibliography
\newpage
\appendix
\section{Appendix}

\subsection{Bayesian logistic regression}
\begin{figure}[htbp]
\centering
\includegraphics[width=0.7\textwidth]{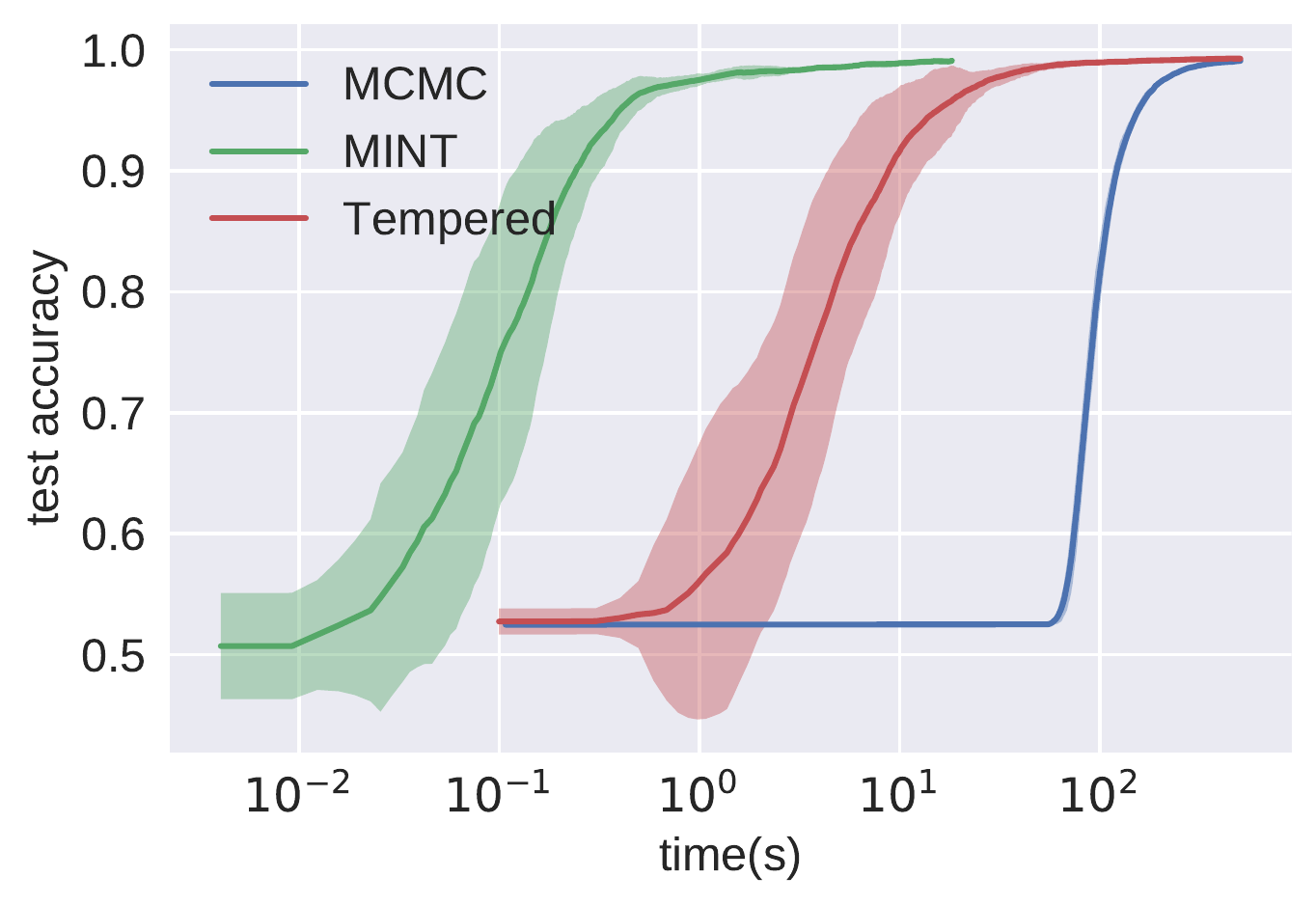}
%\vspace{-2mm}
\caption{ Test accuracy for MINT,  tempered MCMC and MCMC as a function of wall clock time.}
\label{fig:test_time}
\end{figure}

As mentioned in Section 3.2, we include a plot the test accuracy of the three samplers against wall clock time in Figure \ref{fig:test_time}.

\subsection{MINTEE}

In this section we provide details about the MINTEE sampler presented in Section 4. We first review the essentials of the EE sampler \cite{kou2006discussion} and then provide the pseudo-code for MINTEE. We also include the implementation details and more experimental results for the experiment in Section $4.3$ .

\subsubsection{The MINTEE sampler}

%%% How does EE work %%%
As mentioned in Section 4.2, the EE sampler also considers a temperature ladder: $ 1 = T_0 < T_1 < \cdots T_{K-1} <\infty
$. In addition, it considers an increasing energy sequence:
\begin{equation}\label{energy_seq}
	H_0<H_1<H_2<\cdots H_{K-1}< H_{K} = \infty
\end{equation}

where $H_0$ is a lower bound for the energy function: $H_0\leq\inf_xh(x)$. This defines $K$ distributions, each indexed by a temperature and an energy truncation: $\pi_k(\theta) \propto \exp{-\frac{h(\theta)\vee H_k}{T_k}}$ for $k=0,1,\cdots K-1$. For each $k$, a sampling chain targeting $\pi_k(\cdot)$ is constructed.  Clearly, $\pi_0(\cdot)$ is the original target distribution. The EE sampler employs the other $K$ chains to overcome local trapping by incorporating a type of transition called the ``Equi-Energy  jump (EE jump)'' to exchange states between two adjacent chains.

The idea is to first partition the state space $\Theta$ based on the energy levels: $\Theta=\cup_{k=0}^{K-1}D_k$, where $D_k =\{\theta:h(x)\in[H_k,H_{k+1}) \}, 0\leq k \leq K-1$ are the energy sets determined by the predetermined energy sequence \eqref{energy_seq}. The empirical Equi-Energy  sets are referred to as energy rings. For any $\theta\in\Theta$, let $I(\theta)$ denote the partition index such that $I(\theta) = j$ if and only if $\theta\in D_j$, or equivalently, if $h(\theta)\in[H_j, H_{j+1})$. We start the EE sampler by running a Markov chain $\theta^{(K-1)}$ targeting the highest-order distribution $\pi_{K-1}(\cdot)$. After an initial burn-in period, the EE sampler starts constructing the ${K-1}^{th}$ order energy rings $\hat{D}_j^{(K-1)}, 0\leq j\leq K-1$. The chain $\theta^{(K-2)}$ is updated through two types of transitions: the local MH move and the EE jump. At each update, with probability $p_{ee}$ the current state $\theta_i^{(K-1)}$ goes through an EE jump: a state $\theta^\prime$ is chosen uniformly from the highest-order energy ring $\hat{D}_j^{(K-1)}$ indexed by $J = I(\theta_n^{(K-2)})$ that corresponds to the energy level $\theta_i^{(K-2)}$. Note $\theta^\prime$ and $\theta^{(K-2)}_i$ have similar energy levels, since $I(\theta^\prime) = I(\theta^{(K-2)}_i) $  by construction. The chosen $\theta^\prime$ is accepted to be the next state $\theta^{(K-2)}_{i+1}$ with probability $\min\{1, \frac{\pi_{K-2}(\theta^\prime)\pi_{K-1}(\theta^{(K-2)}_i)}{\pi_{K-1}(\theta^{(K-2)}_i)\pi_{K}(\theta^\prime)}\}$. If $\theta^\prime$ is not accepted, $\theta_{i+1}^{(K-2)}$ keeps the old value $\theta_{i}^{(K-2)}$. After a burn-in period on $\theta^{(K-2)}$, the EE sampler starts the construction of the second highest-order energy rings $\hat{D}_j^{(K-2)}$ in much the same way as the construction of  $\hat{D}_j^{(K-1)}$, that is, collecting the samples according to their energy levels. Once the chain $\theta^{(K-2)}$ has been running for $N$ steps, the EE sampler starts $\theta^{(K-3)}$ targeting $\pi_{K-3}(\cdot)$ while it keeps on running chain $\theta^{(K-2)}$ and $\theta^{(K-1)}$. Like $\theta^{(K-2)}$, at each step chain $\theta^{(K-3)}$ is updated through either a MH update or an EE jump. In this way, the EE sampler successively moves down the energy and temperature ladder until it reaches the distribution $\pi_0(\cdot)$. It can be proven, under mild assumptions, each chain $\theta^{(k)},k=0,1,\dots,K-1$ constructed by the EE sampler is ergodic with $\pi_k(\cdot)$ as its steady-state distribution. The proof and more implementation details can be found in \cite{kou2006discussion}.

%%% How does EE compare with parallel tempering %%%

\begin{figure}
\centering
	\includegraphics[width=0.5\textwidth]{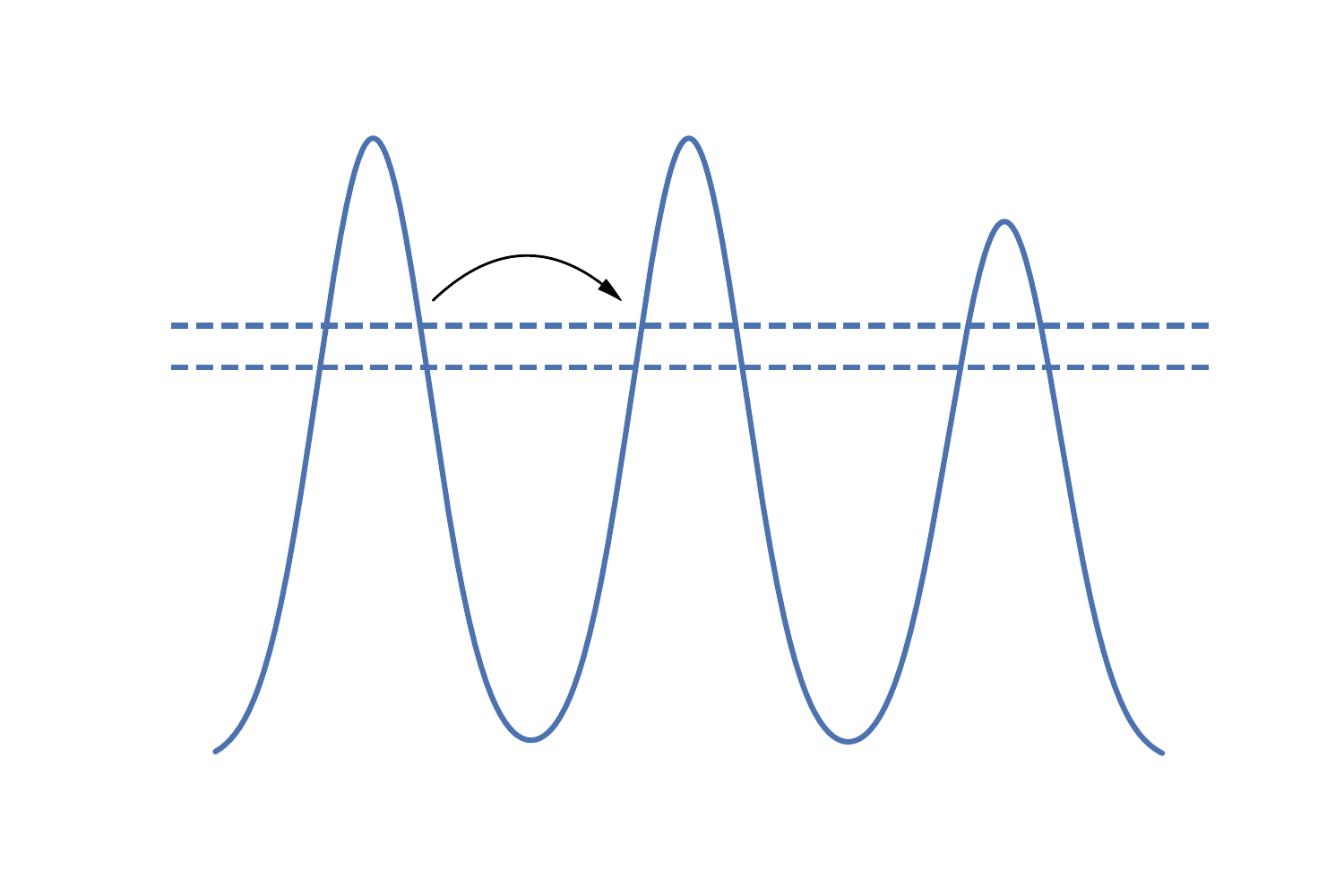}
		%\vspace{-7mm}
	\caption{Illustration of the power of Equi-Energy  jump.}
	\label{fig:eejump}
	%\vspace{-2mm}
\end{figure}

The power of an EE jump is illustrated in Figure \ref{fig:eejump}. As long as the higher temperature chain have previously visited two regions of similar energy values, the EE jump makes it possible for a lower temperature chain to jump across the two regions, even if they are quite far apart in the parameter space and  separated by high energy barriers. In contrast, with parallel tempering, the lower temperature chain can make use of the higher temperature chain only through its current state but not the energy level. Thus the EE sampler makes much stronger use of the information from the higher temperature chain by working with the energy function directly.

%An important difference with the EE sampler and parallel tempering is that is aims to overcome the difficultly of moving across energy barriers by working on the energy function directly. To introduce our method, below we briefly review the EE sampler. Readers are referred to the original paper [kou, zhou and wong 2006] for more details.

\begin{figure*}[htbp]
\centering
	\includegraphics[width=\textwidth]{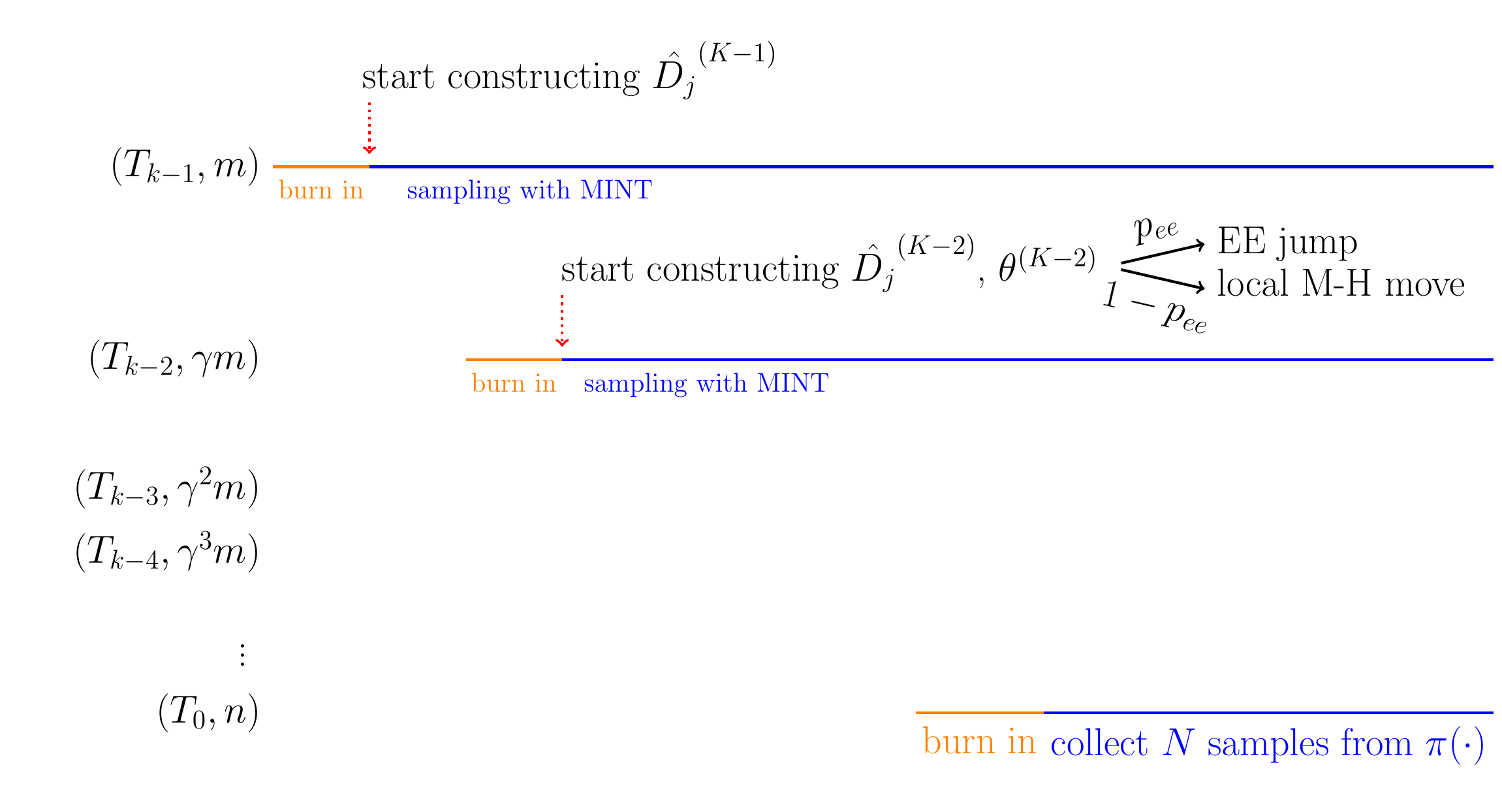}
		
	\caption{Illustration of the MINTEE sampler.}
	\label{fig:ee}\end{figure*}

After introducing the EE sampler, we provide the pseudo code for MINTEE in Algorithm \ref{algo:MINTEE}.  An pictorial illustration of MINTEE can be found in Figure \ref{fig:ee}.   In our implementation, we cached the past states together with their energy values for each chain. We use the cached data from higher temperature chains to construct the Equi-Energy  rings.

Next we address a unique issue caused by mini-batching. Consider two adjacent chains in the temperature ladder. In Type I moves, the chains are updated independently using mini-batch estimates of $\mu(\theta)$. The batch size in a lower temperature chain is larger than that in a higher temperature chain. When we try to decide whether to exchange state $\theta$ in the lower temperature chain with state $\theta^\prime$ in the higher temperature chain, we need to know $\mu(\theta)$ and $\mu(\theta^\prime)$ up to the same required accuracy (which is implicitly determined by the batch size). However, $\mu(\theta^\prime)$ is estimated using a smaller batch size than what is needed for energy evaluation in the lower temperature chain. Therefore, we must draw a further batch of size $m_{k} - m_{k+1}$ to make up for the difference in batch sizes. The extra computational cost for this exchange operation is quite modest. 

\subsubsection{Implementation details of the MINTEE experiment}
\begin{table*}[!htbp]
  \caption{Percentage of samples in each energy ring}
  \label{tab:MINTEE:50d}
  \centering
  \begin{tabular}{c|c|c|ccccccc}
    \toprule
 \multicolumn{3}{c|}{} & \multicolumn{7}{c}{Energy ring sizes ($\%$) }\\
    \midrule
   Chain &  $T_k$ & $m_k$ & $<H_1$ & $[H_1, H_2)$ & $[H_2, H_3)$ & $[H_3, H_4)$
    & $[H_4, H_5)$    & $[H_5, H_6)$   & $\geq H_6$\\
    \midrule
$\theta^{(0)} $ & 1.00 & 100000  & 51.08 & 4.76 & 6.45 & 8.44 & 10.00 & 9.73 & 9.55\\
$\theta^{(1)} $ & 19.41 & 5378  & 40.15 & 5.01 & 7.14 & 9.63 & 12.20 & 12.72 & 13.16\\
$\theta^{(2)} $ & 27.13 & 3841  & 34.80 & 4.91 & 7.16 & 10.11 & 13.09 & 14.37 & 15.57\\
$\theta^{(3)} $ & 37.93 & 2743  & 29.93 & 4.78 & 7.05 & 10.23 & 13.97 & 15.82 & 18.22\\
$\theta^{(4)} $ & 53.02 & 1959  & 22.20 & 4.17 & 6.42 & 9.93 & 14.40 & 18.17 & 24.71\\
$\theta^{(5)} $ & 74.06 & 1400  & 13.60 & 3.16 & 5.12 & 8.45 & 13.69 & 20.02 & 35.96\\
$\theta^{(6)} $ & 103.51 & 1000  & 6.08 & 1.74 & 3.07 & 5.60 & 10.44 & 18.71 & 54.37\\
    \bottomrule
  \end{tabular}
\end{table*}

Recall the model we consider is as follows, $x\in\reals$:
\begin{equation*}\label{eq:50d}
	x\sim \frac{1}{50}\Gsn(\theta_1, \sigma^2) + \frac{1}{50}\Gsn(\theta_2, \sigma^2) + \cdots + \frac{1}{50}\Gsn(\theta_{50}, \sigma^2)
\end{equation*} where $\sigma^2=1$ is assumed to be known;                          $\theta=(\theta_1, \theta_2,\cdots,\theta_{50})\in\reals^{50}$ is the parameter of interests. We generate $10^5$ observations from $\theta^* = [2, 0, \cdots, 0]$. Due to the symmetry of the model in $\theta$, the posterior has 50 modes at $2e_j$, $j = 1,\cdots, 50$, where $e_j$ is a standard basis vector in $\reals^{50}$. Without lose of generality, we put a uniform prior over $\theta$. We apply MINTEE to this problem. We take $H_0$ to be $-\sum_{i=1}^n\log p(x_i|\theta^*)$, although any of the 50 modes in this case has the same lowest energy level. We set $K$  to be $7$. In other words $7$ chains were employed in total. We increase the batch size by $1.4$ every time we start a new chain. Note the temperature sequence is determined once the batch size and its rate of increase is chosen. Following the suggestion by \cite{kou2006discussion}, we set energy levels $H_1, \cdots, H_6$ by letting $(H_{k+1} - H_k ) / T_k \approx c=10$. The EE jump probability $p_{ee}$  was taken to be $0.1$. More details about the  configurations of each chain are tabulated in Table \ref{tab:MINTEE:50d}. The initial states of all  chains were taken to be the origin. In order to make use of the gradient information of the log posterior, we use Langevin dynamics to generate proposal. To be specific, for each chain $k$, a proposal is generated as:
\begin{equation*}
	\theta^{(k)\prime }= \theta^{(k)} + \frac{\epsilon_{k}^2}{2} g^{(k)} + \epsilon_{k}z
\end{equation*} where $z\sim\Gsn(0,I_{50})$; $g^{(k)}$ denotes the stochastic gradient of $\log {\pi_k}(\theta) = -\frac{h(\theta)\vee H_k}{T_k}$  evaluated at $\theta^{(k)}$. The step size $\epsilon_k$ of each chain is initialized as $5e^{-4}\sqrt{T_k}$ and then adaptively chosen so that the acceptance probability of each chain is within $[0.2, 0.5]$. Each chain starts to construct its energy ring after a burn-in of 10,000 samples. We run the algorithm to collect 300,000 samples from the lowest temperature chain. We report the percentage of samples in each energy ring in Table \ref{tab:MINTEE:50d}. One can see that as the temperature decreases, samples become more and more concentrated in the low energy rings.

%\begin{figure}
%\centering
%	\includegraphics[width=0.5\textwidth]{langevin_50d_3e5_1d.pdf}
%	\caption{1-D histogram of MINTEE's samples from the true posterior distribution.}
%	\label{fig:mintee_dist}
%\end{figure}

\begin{algorithm*}[htbp]                   % enter the algorithm environment
\caption{MINTEE}
\label{algo:MINTEE}                           % and a label for \ref{} commands later in the document
\begin{algorithmic}[1]                    
	\REQUIRE $N, B$ (number of burn-in samples)
    \ENSURE Samples from $K+1$ chains: $\theta_i^{(k)}\sim\pi_k(\theta)$, $k=0, 1, \dots, K$, $i=1,2,\dots$
    \STATE $n = 0$
    \WHILE{number of samples in chain 0 is less than $N$}
    	\STATE $n = n + 1$
    	\FOR{$k = K$ down to $0$}
    		\IF{$n < (K-k)(B+N)$}
    			\STATE continue
    		\ENDIF
    		\STATE Retrieve $(\theta^{(k)}_{n-1}, \hat{h}(\theta^{(k)}_{n-1}))$  from cached results
    		
    		\STATE \COMMENT{$\hat{h}(\theta^{(k)})$ denotes the estimated energy of $\theta^{(k)}$ based on batch size $m_{k}$ }	    
    		
    		\STATE Compute $i = I(\theta^{(k)}_{n-1})$
    		\STATE Draw $u_{mh}$ uniformly from $[0,1]$
    	    \IF{$k==K$ or $\hat{D}^{(k+1)}_i=\emptyset$ or $u_{mh} < 1-p_{ee}$}
    	    	\STATE perform an MH step on $\theta^{(k)}_{n-1}$ to get $\theta^{(k)}_n$
    	    \ELSE 
    	    	\STATE \COMMENT{perform an EE jump on $\theta^{(k)}_{n-1}$}
    	    	\STATE Uniformly pick a state $(\theta^\prime, \hat{h}(\theta^\prime))$ from $\hat{D}^{(k+1)}_{i}$ 
   
    	    	\STATE Increase the accuracy of $\hat{h}(\theta^\prime)$ by sampling additional $m_{k} - m_{k+1}$ samples
    	    	
    	    	\STATE Draw $u_{ee}$ uniformly from $[0,1]$
    	    	
    	    	\IF{$u_{ee}<\min\{1, \frac{\hat\pi_k(\theta^\prime)\pi_{k+1}(\hat\theta^{(k)}_{n-1})}{\hat\pi_{k+1}(\theta^\prime)\hat\pi_{k}(\theta^{(k)}_{n-1})}\}$}
    	    	\STATE \COMMENT{Here $\hat{\pi}_k(\theta) = \exp{-\frac{\hat{h}_k(\theta)\vee H_k}{T_k}}$, where $\hat{h}_k(\theta)$ denotes the estimated energy of $\theta$ based on a batch size of $m_k$}
    	    	\STATE $\theta^{(k)}_n = \theta^\prime$
    	    	\ELSE
    	    	\STATE $\theta^{(k)}_n = \theta^{(k)}_{n-1}$
    	    	\ENDIF
    	    \ENDIF
    	    \IF{$n>(K-k)(B+N)+B$}
    	    	\STATE Add $(\theta^{(k)}_n, \hat{h}(\theta^{(k)}_n))$ to $\hat{D}^{(k)}_{I(\theta^{(k)}_n)}$
    	    \ENDIF
    	\ENDFOR
    \ENDWHILE
\end{algorithmic}
\end{algorithm*}

\end{document}